# Selecting Classifiers by Pooling over Cross-Validation Results in More Consistency in Small-Sample Classification of Atrial Flutter Localization

Muhammad Haziq Kamarul Azman[1], Olivier Meste[2,3], Kushsairy Kadir[1]

[1]British Malaysian Institute, Universiti Kuala Lumpur, Gombak, Malaysia
[2]Laboratoire I3S, Sophia-Antipolis, France
[3]Université Côte d'Azur, Nice, France

**Abstract**

*Selecting learning machines such as classifiers is an important task when using AI in the clinic. K-fold cross-validation is a practical technique that allows simple inference of such machines. However, the recipe generates many models and does not provide a means to determine the best one. In this paper, a modified recipe is presented, that generates more consistent machines with similar on-average performance, but less extra-sample loss variance and less feature bias. A use case is provided by applying the recipe onto the atrial flutter localization problem.*

## 1. Introduction

Artificial intelligence (AI) systems in the healthcare industry are increasingly common in recent years due to the massive development in computation capacity and accessibility. Of note, is the employment of AI in the clinic for the aid of clinical diagnosis [1].

Learning algorithms constitute a part of the process by which AI systems infer knowledge from the available data. It is a challenge to obtain good learning machines that balances between the assumptions made inside the algorithm, and the knowledge provided by the input data. This is aggravated when data is scarce and there are many input features, and is notably true in the clinical setting (e.g. rare diseases characterized using genomic data).

In practice, learning machines are inferred through a process of sample re-use [2]. One such process discussed in this paper is the *K*-fold cross-validation (KFCV). KFCV allows practical inference of learning machines and works well in small-sample situation.

In a typical KFCV recipe, the dataset is split into $K$ equal partitions or fold, and $K$ models are inferred from the union of $K$-1 leftovers. When selecting a machine, the fold is usually selected first (usually the fold with the smallest error), and then the feature subset.

However, because KFCV produces, by design, $K$ different machines, it is a question how one must select the best machine and the best feature subset. Even if the selection procedure is reversed (i.e. select the best feature first, then the best fold), one will have to determine how the features have to be selected over the different folds. No formal methodology exists, to the authors' knowledge.

In this paper, an original methodology is proposed that adds to the original KFCV recipe and allows one to estimate the best learning machine, based on a selection of the best feature subset using pooled metrics, and then the best fold. This makes the problem of feature selection independent of any individual fold. In return, the modified KFCV recipe returns a learning machine with a more consistent performance as opposed to the original KFCV recipe.

## 2. Materials and Method

All processing and learning steps and arrangements are described below. All work was done using MATLAB (Mathworks, USA).

### 2.1. Context of the research

Atrial flutter (AFL) is a disease where the atrium continuously and regularly depolarizes due to a perpetual rotating activation wave (see Figure 1 (a)). They can occur in either the right or left atrium. AFL can be cured by ablation therapy, which breaks the conduction circuit inside the atrium and stops the rotating wave. Classifying AFL localization (right or left) helps in improving cost of ablation procedure [3]. It has been shown previously that localization using variability in the repetitive flutter waves allow good localization of AFL prior to ablation [4].

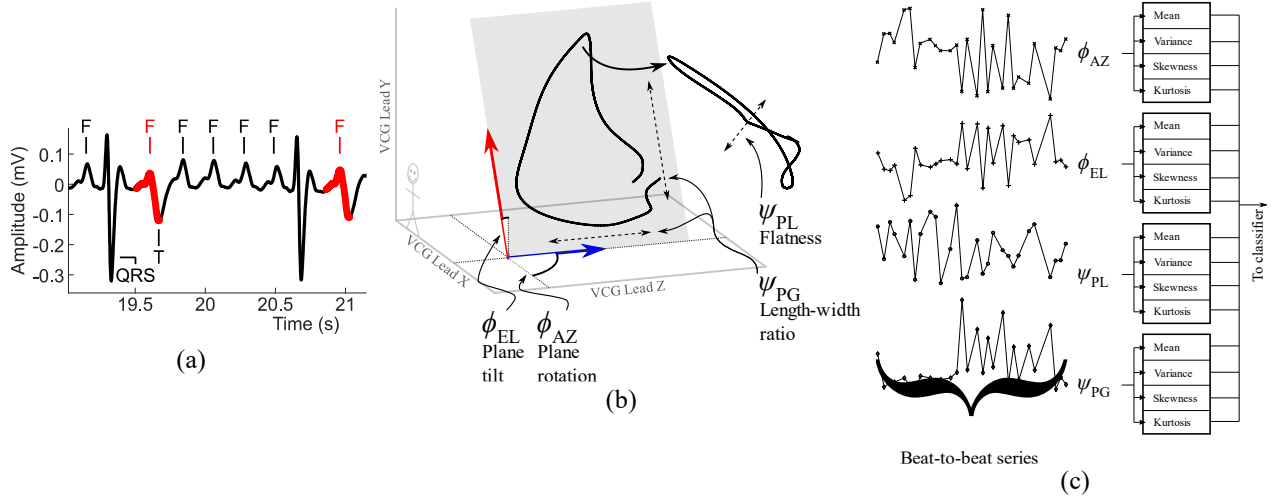

Figure 1. Flow of data processing. (a) ECG signal from a single lead with AFL, with F waves marked. (b) Characterization of AFL VCG loops using several parameters. (c) Calculation of variability features from the beat-to-beat series.

## 2.1. Dataset preparation

The main raw data used in this paper were 75 standard 12-lead digital ECG recordings of atrial flutter, each of about 1 minute in duration. These were acquired from a recording database at Centre Hospitalier Princesse Grace, Monaco with permission. Recordings had been taken during catheter ablation procedure, and the right or left localization was determined for each recording based on the procedure report. After selection, a total of 56 recordings (31 right, 25 left) were used.

The general processing of these signals has been described in detail in [4]. The objective of this was to extract the variability in each F wave and use it to classify right or left localization. Briefly, the signals were filtered to remove motion artifacts, powerline noise and other high-frequency noises. F waves were detected using an original detection algorithm and segmented. The number of waves obtained per record was 64 ± 48 (mean ± standard deviation).

Each F wave was transformed into a vectorcardiogram (VCG) [5]. The vectorcardiogram is a three-lead system based on a linear combination of the 12-lead ECG. The VCG system is aligned to the anatomical axis (lead X: left-right, lead Y: up-down, lead Z: front-back). The F waves become a loop in this system. From this loop, several characteristic parameters were calculated such as its orientation, flatness and circularity (see Figure 1 (b)).

The result of this step is the obtention of several beat-to-beat series of VCG loop parameter values, clean of any source of artificial variability, and reflecting the beat-to-beat changes of the AFL circuit. Variability can be measured from this series by calculating the following higher-order statistics: 1) the mean, 2) the variance, 3) the skewness and 4) the kurtosis, for all 4 loop parameters, for all 56 recordings. In total, there were 16 variability features of AFL (see Figure 1 (c)).

## 2.2. Learning setup

The initial dataset comprises of 56 data samples. This initial set is then split using a stratified holdout scheme, with 30% of the data (16 data samples) held out for test. The remaining 40 data samples are then split using a stratified 5-fold cross-validation scheme. Each fold contains 8 data samples. One fold will be left out as the validation set, whereas the remaining four folds become the training set.

Due to the limited amount of training data (ideally 16 data samples per class), it is not reasonable to use advanced classifiers such as neural networks or trees. Therefore, in this paper, a linear classifier is considered. The support vector machine (SVM) classifier with linear kernel was used as the learning machine, with a box constraint set to 1.

The objective of the KFCV recipe is to find the best fold and feature subset. The evaluation metric used here is the zero-one loss. The selection is commonly performed by taking the best fold for a fixed subset of features. The question of which fold to select is easy: take the one with the smallest validation loss. However, this does not guarantee the optimality of the feature subset as it is related to the fold (i.e. a function of the fold).

Instead, the validation losses should be pooled over all the $K$ folds. This allows us to obtain a single loss value for each feature subset. Then, it is easy to select the best feature subset (e.g. taking the one with the smallest number of features and the smallest loss). Selection of the best fold is then made as usual. The modified recipe can be seen in Figure 2. Note that in the classic KFCV recipe, the best fold is selected first, before the feature.

Once the best model is selected, the held-out portion of the initial dataset is used to test the classifier performance and to obtain the final performance metric. To ensure the consistency of the comparison, the recipe is repeated for 30 iterations. In each iteration, the initial dataset is randomly split again, as with the cross-validation split.

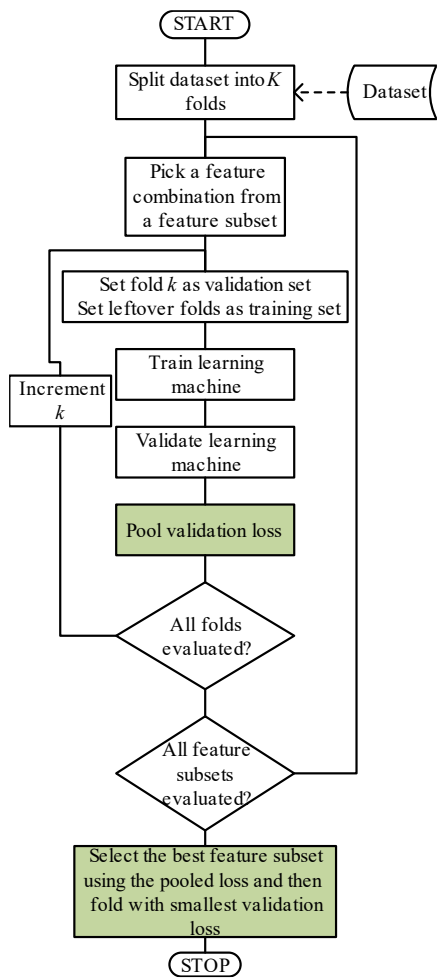

Figure 2. Modified KFCV algorithm for model selection. The green boxes represent the original contribution of this paper.

## 3. Results

For the purpose of comparison, the common portions of the recipe, as detailed in Figure 2 in white boxes, is performed in tandem for the two recipes. This way, the randomness in data splitting and model training is avoided. The selection process technically differs only by the pooling of the validation loss and the selection criteria.

### 3.1. Analysis of extra-sample loss

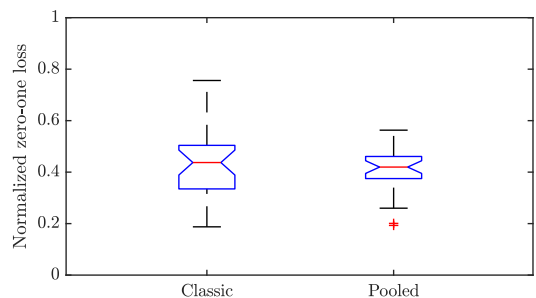

Figure 3. Boxplot of normalized zero-one loss for each recipe.

Figure 3 illustrates the normalized zero-one loss, obtained from applying the best classifier on the test set, for 30 repetitions. Table 1 summarizes the statistics of the repetitions. It can be seen that the median loss is similar in both cases (0.44 vs. 0.42, classic vs. pooled, $p$=NS using Mann-Whitney $U$ test), indicating that the two recipes produce similar on-average performance. However, it can be seen that the spread of the loss is smaller in the case of pooled KFCV (interquartile range 0.17 vs. 0.09, classic vs. pooled). This suggests that the process returns classifiers with a consistently similar performance.

Table 1. Summary statistics of the test loss

|  | Classic | Pooled |
|---|---|---|
| Max | 0.76 | 0.56 |
| 75th quantile | 0.50 | 0.46 |
| Median | 0.44 | 0.42 |
| 25th quantile | 0.33 | 0.38 |
| Min | 0.19 | 0.19 |
| Mean | 0.44 | 0.42 |
| Interquantile range | 0.17 | 0.09 |
| Standard deviation | 0.13 | 0.10 |

### 3.2. Feature selection bias

The best feature subset is analyzed. It was found that over the 30 repetitions, the classic recipe produces 24 unique subsets. The pooled recipe produces only 18 unique subsets. This shows the classifier consistency when using pooled KFCV.

Each of the 16 features are assigned a feature score, defined as the number of times the feature was used over the entire unique set of feature subsets, divided by the number of unique feature subsets. This measure illustrates partly the relevance of the feature in the classification. This strategy is similar to the one employed in [4] but not equivalent. The scores are shown in Table 2. The all train score shown is for the linear SVM.

Table 2. Feature score on unique feature subsets

| Features | Feature score | | |
|---|---|---|---|
| | Classic | Pooled | All train* [4] |
| Mean($\phi_{AZ}$) | 0.13 | 0 | 0.50 |
| Variance($\phi_{AZ}$) | 0.21 | 0.11 | 0.56 |
| Skewness($\phi_{AZ}$) | 0.25 | 0.11 | 0.63 |
| Kurtosis($\phi_{AZ}$) | 0.13 | 0.11 | 0.56 |
| Mean($\phi_{EL}$) | 0.04 | 0.06 | 0.63 |
| Variance($\phi_{EL}$) | 0.04 | 0.11 | 0.75 |
| Skewness($\phi_{EL}$) | 0.04 | 0.17 | 0.50 |
| Kurtosis($\phi_{EL}$) | 0.13 | 0 | 0.69 |
| Mean($\psi_{PL}$) | 0 | 0 | **0.81** |
| Variance($\psi_{PL}$) | 0 | 0.06 | 0.69 |
| Skewness($\psi_{PL}$) | 0.29 | 0.28 | **0.94** |
| Kurtosis($\psi_{PL}$) | 0.04 | 0.06 | **0.88** |
| Mean($\psi_{PG}$) | 0.17 | 0.33 | **0.94** |
| Variance($\psi_{PG}$) | 0.04 | 0.11 | 0.19 |
| Skewness($\psi_{PG}$) | 0.29 | 0.39 | 0.50 |
| Kurtosis($\psi_{PG}$) | 0 | 0.06 | 0.56 |

*Feature score is calculated differently, and on the whole initial dataset. Bold indicates relevant features.*

An arbitrary cutoff of 0.8 was used to determine relevance, based on the all train feature score. Four features are thus found to be relevant. Comparing the feature scores calculated from the best subsets from the two KFCV recipes, it is seen that two relevant features, Skewness($\psi_{PL}$) and Mean($\psi_{PG}$), are found to tally well with the all train score. This shows that both recipes can find the relevant features from the dataset. Two other relevant features, Mean($\psi_{PL}$) and Kurtosis($\psi_{PL}$), are both found irrelevant by the feature score. Upon investigation, it is found that these two features were relevant when considering feature subsets of size three and above; the current paper limits the subset up to size two.

From the data in Table 2, it can be seen that when the feature is irrelevant, classic KFCV tends to overemphasize the feature, as opposed to pooled KFCV. This may be explained due to the spurious nature of the selection process that bases its initial selection on folds rather than on features.

## 4. Discussion and Conclusion

It is crucial, in employing AI in the clinical diagnosis, to obtain learning machines that have good performance. In practice this means a careful selection of the training data as well as the features it must use. Although *K*-fold cross-validation is a popular and effective technique for constructing such machines, it has been shown that the standard recipe has some downsides.

Pooling has been suggested in a reference machine learning text [2], although not explicitly demonstrated. Here, this paper implements by means of a practical recipe, and demonstrates the concept along with a comparison. It is worthy of note that the recipe here is not exclusive to atrial flutter classification alone, but is applicable in all similar situations, regardless of the field of study.

## Acknowledgments

The first author would like to thank Dr. D. G. Laţcu from Centre Hospitalier Princesse Grace for providing the clinical data used in this research.

Address for correspondence:

Muhammad Haziq Kamarul Azman
Electrical Engineering Section
Universiti Kuala Lumpur, British Malaysian Institute
Jalan Sg Pusu, Batu 8,
53100 Gombak
Selangor Darul Ehsan
Malaysia